\documentclass[]{raa}                  
\usepackage{graphicx,times}             
\input{epsf.sty}                        
\input{psfig.sty}                       

\begin{document}

   \title{Diagnostic functions of solar coronal magnetic fields from radio observations}

   \volnopage{Vol.00 (2022) No.0, 000--000}      
   \setcounter{page}{1}          

   \author{Baolin Tan \inst{1, 2}}
   \institute{CAS Key Laboratory of Solar Activity, National Astronomical Observatories of Chinese Academy of Sciences, Beijing 100012, China, {\it bltan@nao.cas.cn}\\
   \and School of Astronomy and Space Science, University of Chinese Academy of Sciences, Beijing 100049, China\\ }

   \date{Received~~2022 month day; accepted~~2022~~month day}

\abstract{In solar physics, it is a big challenge to measure the magnetic fields directly from observations in the upper solar atmosphere, including the chromosphere and corona. Radio observations are regarded as the most feasible approach to diagnose the magnetic field in solar chromosphere and corona. However, because of the complexity and diversity of the emission mechanisms, the previous studies have only presented the implicit diagnostic functions of the magnetic field for specific mechanism from solar radio observations. This work collected and sorted out all methods for diagnosing coronal magnetic field from solar radio observations, which are expressed as a set of explicit diagnostic functions. In particular, this work supplemented some important diagnostic methods missed in other reviews. This set of diagnostic functions can completely cover all regions of the solar chromosphere and corona, including the quiet region, active region and flaring source regions. At the same time, it also includes incoherent radiation such as bremsstrahlung emission of thermal plasma above the quiet region, cyclotron and gyro-synchrotron emissions of magnetized hot plasma and mildly relativistic nonthermal electrons above the active regions, as well as coherently plasma emission around flaring source regions. Using this set of diagnostic functions and the related broadband spectral solar radio imaging observations, we can derive the magnetic fields of almost all regions in the solar atmosphere,which may help us to make full use of the spectral imaging observations of the new generation solar radio telescopes (such as MUSER, EVOSA and the future FASR, etc.) to study the solar activities, and provide a reliable basis for the prediction of disastrous space weather events.
}

   \authorrunning{Baolin Tan}            
   \titlerunning{Radio diagnostic functions of solar coronal magnetic fields}  

   \maketitle
\section{Introduction}
\label{sect:intro}

In solar physics, there are many important unsolved mysteries, including the coronal heating, the origin of solar eruptions, and the structure formation of solar atmosphere, etc. As for the coronal heating mechanism, so far, more than tens of models are proposed to explain the formation of the hot upper chromosphere and very hot corona, which can be summarized into three classes: (1) Wave-heating (AC mechanism, Davila 1987, De Pontieu et al. 2007, etc.), (2) Magnetic reconnection heating (DC mechanism, Parker 1988, Sturrock 1999, Rappazzo et al. 2008, etc.), and (3) Magnetic-gradient pumping mechanism (Tan 2014, Tan et al. 2020, Tan 2021). Although these models vary widely, one thing they have in common is that they are so closely linked to the solar magnetic fields that many people say there could be no hot corona without magnetic fields. As for the solar eruptions, almost all of the source regions of solar flares, coronal mass ejections (CMEs) and even the plasma jets are located in the solar atmosphere above the photospheric regions with relative strong magnetic fields, they are intrinsically rapid release of magnetic energy in the chromosphere and corona (Hudson 2011, Fletcher 2011, etc.). In fact, magnetic field dominates all the dynamics, structures and evolutionary phenomena of the solar atmosphere, and even particle accelerations and propagations. We may say that magnetic field ultimately controls the evolution of everything in solar system, including the Sun itself and the disastrous space weather events in the solar-terrestrial and the interplanetary space environment. It is because of the magnetic field that the Sun becomes so colorful and so interesting. Therefore, it is a high priority goal for observational solar physics to measure or diagnose the magnetic fields in the solar atmosphere.

Since coronal magnetic fields are so important in solar physics, then, how can we get accurate information about them? So far, there are several possible methods to obtain the information of coronal magnetic fields which can be listed in following:

(1) Zeeman effect

It is well-known that we can directly measure the magnetic fields by using Zeemen splitting effect of spectral lines (Hale 1908). $\triangle\lambda_{B}\approx4.67\times10^{-13}g_{eff}\lambda^{2}B$, here, $\triangle\lambda_{B}$ is split of the spectral line, $g_{eff}$ is the effective Lande factor, $\lambda$ is the wavelength at \AA, $B$ is the magnetic field at Gauss. Despite the great successes of magnetic field measurements in the solar photosphere and low chromosphere by applying this method at optical and infrared wavelengths, because of the very high temperature and relatively weak magnetic field, the reliable measurements of magnetic fields in the high chromosphere and corona have still remained elusive (Casini, White \& Judge 2017). Zeeman splitting observations of infrared coronal forbidden emission lines Fe XIII 1075 nm have been successfully used to deduce the coronal magnetic field with a sensitivity of about 1 G with 20$"$ spatial resolution after 70 minutes of integration (Lin et al. 2004, Liu \& Lin 2008), but this method only works for measuring coronal magnetic fields outside the solar limb. Due to high temperature and low density, the well-developed magnetic field measurements of using Zeeman effect are hardly applicable to the upper chromosphere and corona.

(2) Hanle effect

The Hanle effect (Hanle, 1924) is the modification of the linear polarization of a spectral line scattered by a local magnetic field, may provide strong diagnostics of regions of weak magnetic fields such as the quiet Sun and the corona, where a number of spectral lines with different but complementary sensitivity ranges are present (Ignace et al. 1997). It has been used to determine the field strength and distribution of magnetic structures of the solar prominence (Raouafi et al. 2016). However, the observation of the associated coronal linear polarization is always very difficult and the interpretation depends on very complicated physical processes.

(3) Faraday effect

The magnetic field in the high corona and the solar wind can be obtained from the Faraday effect of linearly polarized celestial radio sources (Spangler 2005): the angle changes of polarization can be expressed as $\triangle \varphi=C\lambda^{2}\int N_{e}B\cdot ds$, here, $C$ is a constant, $N_{e}$ is the plasma density along the line of sight which need to be obtained from other approach. The Faraday effect can be applied to diagnose the magnetic fields in the high corona and interplanetary space, but it meets many difficulties in the chromosphere and low corona (Hatanaka 1956).

(4) Magnetic-field-induced transition method

The magnetic-field-induced radiation originates from atomic transitions where the lifetime of the upper energy level is sensitive to the local, external magnetic field (Beiersdorfer et al. 2003, Li et al. 2015). Recently, based on the theory of magnetic-field-induced transition (MIT), some people proposed to apply the EUV spectral observations to derive the coronal magnetic field and showed very good prospects (Li et al. 2021, Chen et al. 2021).

(5) Coronal seismology

Coronal seismology is mainly concerning the oscillations in coronal loops, interpretation of the oscillations observed in terms of global MHD waves allows to connect the period of the oscillations and the loops length with the magnetic field strength in the loops. Therefore, the coronal seismology can be applied to diagnose coronal magnetic fields (Nakariakov et al. 1999, Nakariakov \& Ofman 2001, De Moortel \& Pascoe 2009, Andries et al. 2009). Recently, many people applied the coronal seismology to obtain the global magnetogram from 1.05 $R_{sun}$ to 1.35 $R_{sun}$ outside the solar disk (Yang et al. 2020, Jess et al. 2016, etc.).

(6) Modeling extrapolations

Dulk \& McLean (1978) proposed an estimation of the averaged magnetic field distribution in corona. The most economical approach is to derive the coronal magnetic fields by using the modeling extrapolations from the observed photospheric magnetographs (Yan \& Sakurai 2000, Metcalf et al. 2008, Jiang \& Feng 2013, Chifu et al. 2015, Zhu \& Wiegelmann 2018, etc.). But the modeling extrapolation can not provide the dynamical and fast variations of the coronal magnetic fields.

(7) Radio diagnostics

Besides the above direct or indirect methods, the solar radio broadband observations are regarded to be the most important approach for diagnosing the magnetic field in the high chromosphere and corona. Essentially, we can see almost everything occurred in solar chromosphere and corona by using radio observations, including the magnetic fields (Dulk 1985, Bastian, Benz, \& Gary 1998) from the face-on to the limb of the solar disk. Radio emissions are very sensitive to both thermal and nonthermal processes in solar atmosphere, including magnetic field and its variations, plasma instabilities, particle accelerations and propagations. Additionally, solar radio observations span frequencies of more than five orders of magnitude, from sub-millimeter waves to hectometer waves, and the emission source region locates from the low chromosphere to the very high corona, and even the interplanetary space, including solar quiet regions, active regions, and flaring eruptive source regions. The emission source involves a wide variety of plasmas, including the dense partial ionized chromospheric plasma and the fully ionized hot dilute coronal plasmas, as well as the energetic nonthermal electrons. Additionally, radio imaging observation can obtain the solar full-disk images with sub-second cadence, much faster than any other approaches and is very useful for providing fast dynamic images of the source regions. For example, the Mingantu Spectral Radioheliograph (MUSER) can capture nearly 600 solar radio full-disk images every 0.2 seconds and every 25 MHz in the frequency range of 0.4-15.0 GHz at frequency intervals of 25 MHz (Yan et al. 2021). Another important instrument is the Expanded Owens Valley Solar Array (EVOSA) which can also provide high-cadence, spatially-resolved spectral imaging observations in the frequency range of 1-18 GHz (Gary et al. 2018, Fleishman et al. 2020, Chen et al. 2020).

The radio emission mechanism depends on the energy of emitting electrons, which may be different in different frequency ranges and in different source regions. In the solar quiet region, the emitting electrons are mainly thermal equilibrium and the radio emission comes from the bremsstrahlung process. In active regions, the radio emission may produce from cyclotron of thermal electrons at low harmonics of the electron gyrofrequency. And in flaring regions, the emission may be contributed from multiple emission mechanisms, including the gyro-synchrotron emission of mildly relativistic electrons, synchrotron emission of nonthermal high-energy electrons, and coherent plasma emission triggered by some plasma instabilities and nonthermal energetic electrons. Due to the diversity of emission mechanisms and the variability in different regions and at different frequency range, the radio observations with high temporal, spatial and spectral resolutions are necessary for obtaining the maps of magnetic fields in chromosphere and corona, respectively.

Many people have already tried to derive the magnetic fields of solar chromosphere and corona from radio observations (Dulk 1985, Gelfreikh et al. 1987, Kundu 1990, Zhou \& Karlicky 1994, Huang 2006, Huang et al. 2013, Wiegelmann et al. 2014, Yan et al. 2020, etc.). There are several review papers elaborated on this topic (Dulk \& McLean 1978, Kundu 1990, Kruger \& Hildebrandt 1993, White 2005, Casini, White, Judge 2017, Alissandrakis \& Gary 2021). However, so far, most of them are fragmented and mainly based on the gyro-synchrotron mechanism, and presented implicit functions of the magnetic field with respect to the parameters of solar radio observations (Dulk \& Marsh 1982, Huang 2006). These implicit functions are not convenient in practice to derive magnetic fields from the solar lower chromosphere to the higher corona from the observations. Even the latest review article has still missed some very important diagnostic methods. As the new generation solar radio telescopes come into operation, such as MUSER, EVOSA and the future FASR cover a very large space from the solar chromosphere to the corona, and has very high temporal, spatial and spectral resolutions (Yan et al. 2009, Yan et al. 2021), it is urgently needed to provide a complete set of explicit diagnosing functions of magnetic field from the radio observations which may cover all regions above solar disk and most frequency range of the radio emissions.

This paper attempts to collect and compile all the methods of radio coronal magnetic field diagnosis so far, and summarizes them in the form of explicit functions, which may cover the full solar disk, including the solar quiet region, active region, and flaring source regions, so that they can be easily applied to the magnetic field diagnosis of all regions of the solar atmosphere. Section 2 will address the criterion of emission mechanism in a given radio frequency range and target region. Section 3 presents the explicit diagnosing functions of magnetic fields from solar radio observations. Finally, some conclusions and discussion of the diagnostic procedure are summarized in Section 4.

\section{Emission mechanisms and the criterions in solar chromosphere and corona}

As we know, solar radio emission comes from a variety of physical processes, including bremsstrahlung, cyclotron, gyro-synchrotron, synchrotron, and coherent plasma emissions. The emission may produce at different frequency range in different altitudes above the same region of the sun's surface, and the emission mechanism may generate from different radiation processes. At the same time, the emission mechanism may be different even at the same frequency range but in different locations in the solar atmosphere (Dulk 1985, Bastian, Benz, \& Gary 1998). Therefore, it is very important to distinguish the solar radio emission mechanism for diagnosing the magnetic fields at the given frequency range and location. According to the physical conditions (including the possible magnetic field strength, plasma density and temperature, thermal or nonthermal electrons, etc.), the diagnostic target area can be divided into three kinds of distinct regions: quiet region, active region, and flaring region. Here, we try to present the main emission mechanism in different regions and their criterions from radio observations.

\subsection{Emission in and above the solar quiet regions}

The basic feature of the atmosphere in and above the Sun's quiet regions is that it has only a weak magnetic field and thermal plasma without nonthermal electrons. This nature implies that the emission is contributed mainly from the bremsstrahlung mechanism (Dulk, 1985). The bremsstrahlung emission is produced when individual electrons are deflected in the coulomb fields of ions. The emission frequency ($f$) is from the plasma cutoff frequency ($f_{pe}\approx9.0n_{e}^{1/2}$) up to $\frac{v}{d}$,

\begin{equation}
f_{pe}<f<\frac{v}{d}.
\end{equation}

Here, $v$ is the velocity of electron and $d$ is the impact parameter (the distance from ion). In thermal plasma, $d$ is in the range from Laudau length ($\lambda_{L}\approx1.67\times10^{-5}T_{e}^{-1}$) to Debye length ($\lambda_{D}\approx69(\frac{T_{e}}{n_{e}})^{1/2}$), and the velocity $v\approx3.89\times10^{3}T_{e}^{1/2}$, $n_{e}$ is the plasma density, and $T_{e}$ is the temperature. The collision is far-encounter when the impact parameter is in the range of $b_{0}<d<\lambda_{D}$ (here, $b_{0}\approx n_{e}^{-1/3}$ is the averaged distance between electrons), and it is close-encounter when the impact parameter is in the range of $\lambda_{L}<d<b_{0}$. Generally, the far-encounters are much more important than the rarely close-encounters, and $\frac{v}{b_{0}}\approx3.89\times10^{3}n_{e}^{1/3}T_{e}^{1/2}$. Therefore, the solar bremsstrahlung emission mainly occurs in the following frequency range,

\begin{equation}
9.0n_{e}^{1/2}<f<3.89\times10^{3}n_{e}^{1/3}T_{e}^{1/2}.
\end{equation}

For example, for the typical parameters in coronal plasmas, $n_{e}\sim10^{16}$ m$^{-3}$, $T_{e}\sim10^{6}$ K, then the bremsstrahlung emission will occur in a very wide frequency range from 900 MHz to 1.2 THz. However, in hot plasma or with nonthermal energetic electrons, the close-encounter will become considerable, the upper limit of frequency of the bremsstrahlung emission may approach to $\frac{v}{\lambda_{L}}$ and even close to $\frac{v}{\lambda_{d}}$ ($\lambda_{d}$ is the de Broglie wavelength of the electron), and this may produce X-ray emission and even hard X-ray emissions. These facts indicate that bremsstrahlung emission has a very wide frequency range. The bremsstrahlung is an incoherent emission process whose brightness temperature ($T_{b}$) does not exceed the thermal temperature of the radiating background plasma ($T_{e}$) (Benz 2009, Nindos 2020).

The emissivity of bremsstrahlung process is proportional to $n_{e}^{2}T_{e}^{-1/2}$, so the plasmas with low temperature and high density are favorable for bremsstrahlung emission (Raulin \& Pacini 2005). Because both impact parameters and electron velocities are distributed continuously over a wide range, the solar bremsstrahlung emission generally appears as broad band continuum with long duration on the observed spectrum, and the peak frequency depends on the emission measure ($\propto n_{e}^{2}$).

In brief, the criteria of bremsstrahlung can be concluded as: $T_{b}\leq T_{e}$, broadband continuum, long duration, weakly or even non-polarization, and relatively weak magnetic fields.

\subsection{Emission in and above the active regions}

Active regions (ARs) are the most prominent manifestation of the large-scale solar magnetic field in the photosphere of the Sun (Shibasaki et al. 2011). The active regions have relatively strong magnetic field (200 -- 2000 G) and large number of hot plasma loops which may occupied an area extended from $10^{4}$ km to more than $10^{5}$ km (Seehafer 1986). As magnetic cyclotron emission is approximately proportional to $n_{e}T_{e}^{\alpha}B^{\beta}$ ($B$ is magnetic field strength, $\alpha>1$, $\beta>1$), therefore, in and above the solar active region, the bremsstrahlung emission tends to be negligible and the emission is dominated by cyclotron of thermal electrons (Bastian et al. 1998, Nindos et al. 2008) which is produced in thin layers around iso-Gauss surfaces where the magnetic field strength is just fitted to the observing frequency equal to a harmonic of the local gyrofrequency (Lee 2007). The emission frequency can be simply expressed as the resonance condition (Melrose 1980, Robinson \& Melrose 1984, Dulk 1985),

\begin{equation}
f=\frac{n(1+\mu\beta cos\alpha cos\theta)}{\gamma}f_{B}=sf_{B}.
\end{equation}

Here, $f_{B}\approx2.8\times10^{6}B$ is the electron gyrofrequency at unit of Hz, here the unit of magnetic field strength $B$ is Gauss (G). The typical value of $B$ is about 100 - 1000 G in solar active region, and the typical frequency $f_{B}$ is in the range from 280 MHz to 2.8 GHz. $n$ is an integer related to the energy of the radiated electrons. $\mu$ is the index of refraction, and $\beta=v/c$. $\gamma$ is the Lorentz factor of electrons, $\alpha$ is the angle between electron velocity and the magnetic field line (pitch angle) and $\theta$ is the angle between the magnetic field line and the line-of-sight. $s=\frac{n(1+\mu\beta cos\alpha cos\theta)}{\gamma}$ is called harmonic number and not an integer. During the period without solar flares in the active region, the radiating electrons are nonrelativistic thermal electrons, for example in the typical coronal plasma in and above the active region, $T_{e}\sim10^{6}$ K, $\beta\sim10^{-2}$, $\gamma\sim1.0$, and $\mu\sim1.0$, then we may have $n\approx s$. The emission mainly occurs at the fundamental frequency ($s=1$) or low harmonics ($s=$ 2 or 3) which is called as cyclotron emission. The emission at fundamental frequency is mainly along the direction of magnetic field, and the low harmonics emission is at the moderate angle with respect to the magnetic field.

Generally, the integer $n$ is approximated as $n\sim\gamma^{3} sin\theta$. Then the harmonic number ($s$) can be expressed as:

\begin{equation}
s=\gamma^{2}sin\theta(1+\mu\beta cos\alpha cos\theta)\approx\gamma^{2}sin\theta.
\end{equation}

With the increase of electron's kinetic energy ($\propto\gamma^{2}$), the harmonic number $s$ will increase rapidly. As for the mildly relativistic electrons, $s>10$, which is called gyrosysnchrotron emission. Although cyclotron and gyrosysnchrotron radiation are emitted at discrete frequencies, they generally give rise to a broadband continuum due to the variations of magnetic field with height (Kakinuma \& Swarup 1962, Fleishman \& Melnikov 2003). Obviously, both cyclotron and gyrosysnchrotron are incoherent emission, which criteria may include: $T_{b}\leq T_{e}$, and long duration continuum (for the continuously varying of $\gamma^{2}sin\theta$) with strong circular polarization and localized only in and above the strong magnetic field regions.

In addition to solar active regions, the solar internetwork magnetic fields in the quiet regions sometimes may reach to kG magnitude (e.g. Wang et al. 1995, Orozco et al. 2007, Zhou et al. 2013, etc.), and magnetic cyclotron radiation also provides considerable contribution to the emission. However, as the internetwork always occupies only a very small fraction area of the solar quiet region, and the existing telescopes have not been able to distinguish them clearly. With the development of the future next generation of solar radio telescopes, such as the SKA (Nindos et al. 2019), it will be possible to identify the magnetic cyclotron emission of the solar network in the same way that we observe in active regions.

\subsection{Emission around the flaring source regions}

Generally, the flaring source region is only a small patch embedded within the solar active region where flare occurs, usually much smaller than the active region. It is always composed of a series of loops with very hot plasma and relatively strong magnetic fields, and containing a great number of nonthermal, collisionless, energetic electrons with power-law distributions. Fast variation, rapid energy-releasing, and nonthermal process are the prominent features of the flaring source region. With such conditions, the emission should have very high brightness temperature, very short lifetime, and rapidly variations. The emission mechanism may include incoherent emission of nonthermal energetic electrons and coherent emission triggered by some plasma instabilities.

(1) Incoherent emission in flaring source regions

The incoherent emission around the flaring source regions should be gyrosysnchrotron emission of the mildly relativistic electrons and synchrotron of the relativistic electrons ($\gamma\gg 1$) over a broad continuum mainly at the instantaneous direction of velocity. The distribution of radiating electrons is power-law. The emission frequency approximates (Dulk \& Marsh 1982),

\begin{equation}
f\approx f_{B}\gamma^{2}sin\theta
\end{equation}

The synchrotron emission is directed strongly at the direction of the instantaneous electron motion and perpendicular to the magnetic field. On the broadband spectral observations, it appears as a broadband continuum spectrum with relatively short lifetime, such as patches, type V bursts, etc. Because the radiating electrons have high energy which is much more energetic than the ambient plasma, their brightness temperature will be higher than the local thermal temperature but usually not exceed 10$^{10}$ K in solar regime.

(2) Coherent emission in flaring source regions

There are mainly two kinds of coherent emissions in flaring source regions: plasma emission (PE) and electron cyclotron maser emission (ECME).

The first coherent emission around the flaring source regions is PE, which is triggered by nonthermal energetic electrons and the related Langmuir turbulence and plasma instabilities. Intrinsically, the plasma emission is a strong wave coupling process between the Langmuir wave and other low-frequency waves. The frequency of plasma emission can be expressed as (Ginzburg \& Zheleznyakov 1958, Zheleznyakov \& Zlotnik 1975, Robinson \& Benz 2000),

\begin{equation}
f= f_{pe}+f_{3}\approx sf_{pe}.
\end{equation}

Here, $f_{pe}\approx9.0n_{e}^{1/2}$ is the plasma frequency. The unit of plasma density $n_{e}$ is $m^{-3}$ and the typical value is from $10^{15}m^{-3}$ to $10^{17}m^{-3}$ in the flaring source regions, and the corresponding typical plasma frequency is from 280 MHz to 2.80 GHz. $s$ is the harmonic number. $f_{3}$ is the frequency of low-frequency waves, such as magnetosonic wave, electron cyclotron wave, whistler wave, and even scattering Langmuir wave. Generally, $f_{3}\ll f_{pe}$ for magnetosonic wave, electron cyclotron wave, and whistler wave, then $s\approx1$, $f\approx f_{pe}$, which is called the fundamental PE. When $f_{3}$ is the scattering Langmuir wave, $f_{3}\approx f_{pe}+f_{B}$, here, $f_{B}\ll f_{pe}$, then $f=2f_{pe}+f_{B}\approx2f_{pe}$, $s\approx2$, the emission is the second harmonic PE. The fundamental PE is always strong polarization in the sense of O-mode in a magnetized plasma, while the second harmonic PE is usually weak polarization.

The coherent PE may produce various kinds of spectral fine structures with very short lifetime, very narrow frequency bandwidth, and superhigh brightness temperature ($T_{b}\gg10^{10}$ K), such as type II burst, type III burst, and the superfine structures overlapping on type IV continuum including Zebra pattern (ZP) structures (Chernov 2010, Tan et al. 2014), fiber bursts (Karlicky et al. 2013, Wan et al. 2021), spike group bursts(Wang et al. 2008, Chernov et al. 2010, Tan 2013, Tang et al. 2021), etc. These spectral fine structures are intrinsically related to the electron gyrofrequency $f_{B}$ which contains the information of magnetic fields (Chernov 2006, 2011).

Generally, the plasma emission is usually only effective with relatively weak magnetic field ($f_{B}<f_{pe}$).

The second coherent emission in flaring source regions is electron cyclotron maser emission (ECME) which is occurred when the magnetic field is relatively strong or the plasma density is relatively low: $f_{B}>f_{pe}$. The second condition for ECME is a population inversion in the electron distribution as compared with equilibrium, such as the loss-cone distribution in magnetic flux tubes (Twiss 1958, Schneider 1959, Wu \& Lee 1979). Many spectral fine structures, such as the spike bursts are possibly generated from ECME mechanism (Fleishman et al. 2003, Tang et al. 2016).

In a word, different emission mechanisms will dominate in different regions of the solar atmosphere, so that radio emission has different characteristics, including emission intensity ($I_{f}$), brightness temperature ($T_{b}$), peak frequency ($f_{peak}$), bandwidth ($\triangle f$), lifetime ($\tau$), degree of polarization ($d_{p}$), frequency drift rate ($\frac{df}{dt}$) and so on. Through the measurement of these parameters, we can identify the emission mechanism in the diagnosing area, and through the physical relationship between the above parameters and the magnetic field, it is possible to establish a definite coronal magnetic field diagnostic method.

\section{Radio diagnosing functions of magnetic fields}

Generally, most solar radio telescopes are designed to receive two components of left- and right-handed circular polarizations (LCP and RCP). According to the magnetoionic theory (Hartree 1931, Appleton 1932), when a beam of electromagnetic wave propagates parallel to the magnetic field, or the X-mode when a beam of electromagnetic wave propagates perpendicular to the magnetic field, both of LCP and RCP have the cutoff frequencies:

\begin{equation}
f_{Rc}=\sqrt{f_{pe}^{2}+\frac{f_{ce}^{2}}{4}}+\frac{f_{ce}}{2}
\end{equation}

\begin{equation}
f_{Lc}=\sqrt{f_{pe}^{2}+\frac{f_{ce}^{2}}{4}}-\frac{f_{ce}}{2}
\end{equation}

By solving the above two equations, we can get: $f_{ce}=f_{Rc}-f_{Lc}$, and $f_{pe}=(f_{Rc}\cdot f_{Lc})^{1/2}$. Then, if we can measure the cut-off frequencies of LCP and RCP from the solar radio observation, we may obtain simultaneously the magnetic field strength and plasma density in the medium of the magnetized plasma:

\begin{equation}
B=357.1(f_{Rc}-f_{Lc})
\end{equation}

\begin{equation}
n_{e}=1.24\times10^{16}f_{Rc}\cdot f_{Lc}.
\end{equation}

Here, the unit of $f_{Rc}$ and $f_{Lc}$ is GHz, $B$ is at Gauss (G), and $n_{e}$ is at $m^{-3}$. This method can be called as low-frequency cut-off diagnostics, and the error of the diagnosing result of this method depends on the measurement accuracy of the cut-off frequency.

Because the cut-off effect is only produced by the propagation process of electromagnetic waves in magnetized plasma, they have nothing to do with the emission mechanisms. In the solar atmosphere, the magnetic field and plasma density of magnetized plasma decrease rapidly with distance from the solar disk. Therefore, the outer atmosphere can almost be regarded as a transparent medium. The diagnostic results given by the above formula almost reflect the real information near the emission source regions. However, so far, no one has seen this method to diagnose the coronal magnetic field,  and this method is always ignored by other reviews. The main reason is that this method requires that the radio telescope should have sufficient frequency bandwidth and frequency resolution, and can be extended beyond the cut-off frequency to ensure the accurate $f_{Rc}$ and $f_{Lc}$. With the new generation of solar radio telescopes, such as MUSER and the future FASR, it is possible to apply the low-frequency cut-off diagnostics to obtain the information of magnetic fields in solar atmosphere.

At the same time, the emission mechanism should be different from different source region, and the corresponding observational results will have different characteristics, which may provide different diagnostic functions to derive the magnetic fields in the solar emission source regions. In the following sections, we will discuss the diagnosing functions of magnetic fields in different region with different emission mechanism.

\subsection{Diagnosing functions of magnetic fields in and above the quiet regions}

The quiet regions cover most of the Sun's surface. In and above the Sun's quiet regions, there is only a weak magnetic field, and the bremsstrahlung emission of thermal electrons is dominated. From the emission mechanism itself, bremsstrahlung has nothing to do with magnetic field. However, if the source region has magnetic field, then the emission is composed of ordinary mode (O-mode) and extraordinary mode (X-mode), the absorption coefficients of O-mode and X-mode are a bit of different, which can be expressed in the following form (Ratcliffe 1959),

\begin{equation}
\kappa_{\sigma}=0.2\frac{n_{e}^{2}}{T^{3/2}(f+\sigma f_{B}cos\theta)^{2}}.
\end{equation}

Here, $\sigma=1$ means the O-mode, and $\sigma=-1$ means the X-mode. $f_{B}$ is the electron gyrofrequency. $\theta$ is the angle between the magnetic field line and the line-of-sight. Supposing the emission source region is optically thin, and $f\gg f_{B}, f_{p}$, ($f_{p}$ is the plasma Langmuir frequency), then the polarization degree ($d_{p}$) can be approximated as,

\begin{equation}
d_{p}=\frac{\kappa_{x}-\kappa_{o}}{\kappa_{x}+\kappa_{o}}\approx 2\frac{f_{B}}{f}cos\theta.
\end{equation}

Because $f_{B}\propto B$, therefore, we may obtain the diagnosing functions of the longitudinal component of magnetic fields from the measurement of polarization degree,

\begin{equation}
B_{l}=Bcos\theta\approx 178.6fd_{p}.
\end{equation}

Here, the unit of observing frequency $f$ is GHz, and the unit of magnetic field $B_{l}$ is Gauss (G). For example, when $f=1.0$ GHz, and the observed polarization degree $r=10\%$, then the longitudinal component magnetic field is about 18 G. At the same time, if we can obtain the azimuth information of the magnetic field with the help of imaging observations at other wavelengths, such as EUV images, we may obtain the total magnetic field strength ($B$).

In general observations, most solar radio telescopes are designed to receive signals of left-handed (L) and right-handed (R) circular polarizations. In such case, the polarization degree can be expressed as,

\begin{equation}
d_{p}=\frac{R-L}{R+L}.
\end{equation}

In practice, the actual emission source tends to be partially optically thick, and the polarization degree depends not only on the magnetic field strength but also on the temperature gradient of the source region (Bogod \& Gelfreikh 1980, Gelfreikh 1990),

\begin{equation}
d_{p}\approx \delta\frac{f_{B}cos\theta}{f}.
\end{equation}

Here, $\delta=-\frac{dlnT}{dlnf}$ is the spectral index of emission electrons. The diagnosing functions of the longitudinal component of magnetic fields is then rewritten into the following form,

\begin{equation}
B_{l}=357.1\frac{f}{\delta}d_{p}.
\end{equation}

As for the optically thin thermal source region, the energy spectral index $\delta=2$ and the above formula reverts to Equation (13). Obviously, the diagnosing accuracy of magnetic fields mainly depends on the measurements of spectral index $\delta$ and the degree of polarization ($d_{p}$).

Additionally, the observed sense of polarizations may provide the information of direction of magnetic field: the right-handed circular polarization ($d_{p}>0$) indicates that the magnetic field is at an acute angle from the line of sight, while the left-handed circular polarization ($d_{p}<0$) indicates an obtuse angle between the magnetic field line and the line of sight.

So far, there is no appropriate method to obtain measurements of the transverse component of magnetic fields in quiet regions. We can only get these information through some indirect approaches, such as modeling extrapolation.

\subsection{Diagnosing functions of magnetic fields in and above the active regions}

As we mentioned in Section 2.2, the emission is dominated by cyclotron and gyro-synchrotron in and above the active regions. In this case, the opacity of gyroresonance emission from hot plasma allows the measurement of magnetic field strength, and the polarization may provide the information of the direction of magnetic fields. When the magnetic field is at the direction pointing towards the observer, the circular polarization is dominantly right-hand circular, and when the field is at the direction pointing away from the observer, the polarization is dominantly left-hand circular (White 2005). Actually, the opacity of gyroresonance emission is a complicated function of the temperature, density, magnetic field strength, and even the wave mode (X-mode or O-mode) and the direction of the magnetic field with respect to the line of sight (Akhmedov et al. 1982, Alissandrakis et al. 2019). Many people derived the spatial resolved magnetic field strength by fitting the microwave spectrum with gyrosynchrotron model at given spatial locations (Chen et al. 2020, etc.). However, these fitting methods are very complicated and have no implicit diagnostic functions for deriving the magnetic fields.

There are several methods which may present implicit diagnostic functions for deriving the magnetic fields in and above the active regions.

(1) Peak frequency diagnostics

For the typical coronal hot plasma in and above the active region, the gyro-synchrotron emission of thermal electrons becomes optically thick at 3 harmonics for O-mode and at 2 harmonics for X-mode. That indicates that the peak frequency ($f_{pk}$) is $3f_{B}$ for O-mode and $2f_{B}$ for X-mode. This may give a diagnostic function of the magnetic field in the source region,

\begin{equation}
B=357.1\frac{f_{pk}}{s}
\end{equation}

The unit of $f_{pk}$ is GHz, and $B$ is Gauss. $s$ is the harmonic number, $s=2$ for X-mode and $s=3$ for O-mode. Here, we can partition the active region into many small grids and find out the peak frequency in each grid, and then derive the magnetic fields by using the above diagnosing functions (Equ. 17). Because the magnetic field will be different at different height above the solar surface, the peak frequency changes accordingly. For example, we may obtain the peak frequencies at foot-point, loop-leg, and loop-top, respectively, and derive the magnetic field accordingly (White 2005). The shortcoming of this method is that, at some cases, both the second and the third harmonics may contribute to the emission, and the result should be somewhere derived between the second and the third harmonic emission (Alissandrakis et al. 1980).

(2) Spectral diagnostics

In high temperature coronal plasma, it is often difficult to determine the harmonic number of cyclotron emission directly, and it is also difficult to extract the exact peak frequency from observations. Therefore, it is hard to use Equation (17) to derive the magnetic field in the active region. Zhou \& Karlicky (1994) obtained another expression,

\begin{equation}
B=(\frac{c^{2}}{k_{B}T_{b}A_{1}}f_{p}^{1.3+0.98\delta}f^{-0.78-0.9\delta}A_{2}^{-2.52-0.08\delta})^{1/(0.52+0.08\delta)}.
\end{equation}

In this expression, $A_{1}=4.24\times10^{1.4+0.3\delta}(sin\theta)^{0.34+0.07\delta}$, $A_{2}=2.8\times10^{6}$. $k_{B}$ is the Boltzman constant. $T_{b}$ is the emission brightness temperature, which can be derived from the observed value emission flux intensity ($F$) and the scale length of the source region. The electron energy spectral index $\delta$ can be derived from the observed emission spectral index ($\alpha$), $\delta=-1.1(\alpha-1.23)$. The scope of application of the above function is $\delta=2-7$, and the corresponding emission spectral index ($\alpha$) is in the range from -0.6 to -5.1. By using the measurement of polarization degree, the value of $sin\theta$ and then the direction of magnetic field can be obtained (Huang \& Nakajima 2002).

Additionally, Equation (18) also contain a variable of $f_{p}$ which depends of the local plasma density, and this should be derived from other methods or some models.

Huang (2008) applied the imaging observations of NoRH at 17 GHz and 34 GHz, and obtained a vector magnetograph. Wang et al. (2015) used this method and the imaging observation of EVOSA and got the whole map of magnetic field in an active region. Zhu et al. (2021) adopted the above method and obtained the evolution of magnetic fields of the coronal loop with respect to the height above the solar surface.

(3) QPP diagnostics

Quasi-periodic pulsations (QPP) are frequently occurred in the solar observations at optical, UV and EUV, soft X-ray, hard X-ray, and radio wavelengths in the solar active regions with and without flares (Aschwanden 1987, Nakariakov et al. 2010, Li et al. 2015, Inglis et al. 2016, Zimovets et al. 2021). As the radio observations may have very high temporal resolutions (down to millisecond of cadence), they may provide the evidence of multiple periods of QPP with period from decades of minutes to milliseconds (Tan et al. 2010). Similar as seismic waves can be used to detect the internal structure of the Earth, QPP can be used to diagnose the physical parameters in the solar source regions, such as the magnetic field, electric currents, plasma density, etc. In solar active regions, radio emissions are always modulated by some MHD oscillations, such as the fast sausage mode or fast kink mode (Abrami 1970, Aschwanden 1987, Nakariakov \& Melnikov 2009, Tan et al. 2010). These QPPs can be used to derive the magnetic field related to the coronal loops in active regions. For example, when the QPP is associated with fast sausage modes, then the magnetic field can be derived,

\begin{equation}
B\approx2.02\times10^{-16}\frac{an_{e}^{1/2}}{P}.
\end{equation}

Here, $P$ is the period of the QPP, $n_{e}$ is the plasma density, $a$ is the radius of the section of the coronal loop. When the radio QPP is associated with fast kink modes, then the magnetic field can be derived,

\begin{equation}
B\approx6.48\times10^{-17}\frac{Ln_{e}^{1/2}}{P}.
\end{equation}

$L$ is the length of the coronal loop. The parameters $a$ and $L$ can be measured from the imaging observations.

Equation (19) and Equation (20) indicate that the QPP diagnostics depends not only on the measurement of oscillation period ($P$), but also on the measurement of the geometric parameters ($a$ and $L$) and plasma density ($n_{e}$) of the coronal loops. The geometric parameters depends on the imaging observations with broadband frequency and high resolutions, and the determination of plasma density is related to the judgment of the emission mechanism and the supplement of imaging observations at other wavelengths, such as multi-wavelength EUV imaging observations, etc.

It is worth noting that radio QPP doesn't always show up in observations. Its diagnosing results can be used to verify the results derived by peak frequency diagnostics and spectral diagnostics. In fact, the results of the above three diagnostic methods can be verified with each other.

The above three kinds of diagnosing methods, peak frequency diagnostics, spectral diagnostics and QPP diagnostics, any of the above three methods is difficult to give the magnetic field information of the whole solar active region, but they need to complement each other to give a complete magnetograph of the active region.

\subsection{Diagnosing functions of magnetic fields around the flaring source regions}

As we mentioned in Section 2.3, the flaring source region is smaller than the active region and much smaller than the quiet region but with rapid variations. The emission around the flaring source region should be including incoherent synchrotron of the relativistic electrons and the coherent plasma emission. Therefore the diagnostics can also be classified into two kinds.

(1) Spectral diagnostics

When the emission is generated from the incoherent synchrotron mechanism and form various kinds of broadband continuum bursts, such as the bright patches, etc, the magnetic field can be derived from the following diagnosing function (Zhou \& Karlicky 1994),

\begin{equation}
B\approx(\frac{c^{2}}{k_{B}T_{b}f^{2}A_{2}^{5/2}A_{3}})^{2}.
\end{equation}

Here, $A_{3}=2.245\times10^{15}(sin\theta)^{1/2}(\frac{2f}{f_{pk}})^{\frac{\delta-1}{2}}f_{pk}^{-\frac{5}{2}}$. The above function is applicable when the index $\delta$ is in the range of 2 - 6. The diagnosing results depends on the measurements of several spectral parameters, including the peak frequency $f_{pk}$, plasma frequency $f_{p}$, index $\delta$, and flux intensity $F$. Some of these parameters can be extracted only from the broadband spectral observations with high temporal and spectral resolutions.

(2) Diagnostics of spectral fine structures

Many spectral fine structures are associated with solar flares, their formations are related to the magnetic fields in the source region and therefore can be used to derive magnetic fields.

\textbf{Zebra Pattern}

The solar radio Zebra patterns are a kind of spectral fine structures on the solar radio broadband continuum spectrogram (Elgaroy 1959, Slottje 1972, Chernov et al. 2003, Altyntsev et al. 2005, etc.), which consist of several or decades of almost parallel and equidistant stripes, and frequently observed in all phases of solar flares, see the statistics of Tan et al. (2014). Because the structure characteristics of ZP contains the information of magnetic fields in the source region, it is an important tool to diagnose the magnetic fields in the flaring source regions.

When the emission is generated from the coherent plasma emission, then we can apply various spectral fine structures to diagnose the magnetic field in the flaring source region, among which the radio ZP structure is the most suitable for diagnosing the magnetic field for their frequency separation between the adjacent ZP stripes is proportional to the magnetic field strength in the source region. We may approximately express the relationship between frequency separations of ZP stripes and the magnetic fields in the ZP source region,

\begin{equation}
B\approx357.1M_{\alpha}\cdot\triangle f.
\end{equation}

Here, $\triangle f$ is the frequency separation between the adjacent Zebra stripes at unit of GHz. $M_{\alpha}$ is a factor which depends on the formation mechanism of the ZP structure. According to the physical classification (Tan et al. 2014), different type of ZP may generates from different coupling processes of the plasma emission and therefore have different value of $M_{\alpha}$. It is necessary to identify the coupling processes related to the ZP structure before deriving the magnetic fields in the source regions:

As for the Bernstein wave coupling processes, all stripes in a ZP structure are proposed to produce from a small compact source region, the emission originates from nonlinear coupling processes either between two Bernstein waves or a Bernstein wave and and the plasma electrostatic upper hybrid wave. The energetic electrons with non-equilibrium distribution excite longitudinal electrostatic waves at frequency of the sum of Bernstein mode frequency $sf_{ce}$ and the upper hybrid frequency $f_{uh}$: $f=f_{uh}+sf_{ce}$. Here, $s$ is the harmonic number. This model predicts that the frequency separation between the adjacent Zebra stripes should be equal to the electron gyrofrequency $\triangle f=f_{ce}$, and then $M_{\alpha}=1$ (Rosenberg 1972, Zheleznyakov \& Zlotnik 1975, Zaitsev \& Stepanov 1983). If the source region is small enough, the $\triangle f$ should be a constant, and this point is consistent with the class of Equidistant ZP (EZP) (Tan et al. 2014).

As for the whistler wave coupling processes of ZP formation, the whistler waves are generated at the anomalous Doppler cyclotron resonance ($f_{w}-\frac{k_{\|}}{2\pi}+f_{ce}=0$) under larger angles to the magnetic field, they may form standing wave packets in front of the shock wave, and when the group velocity of whistler waves is equal to the shock velocity, a ZP structure with slow oscillating frequency drift will produce. Here, each Zebra stripe corresponds to a propagating whistler wave packet. The emission frequency is $f=f_{pe}+sf_{w}$, here $f_{w}\approx0.1-0.5f_{ce}$ is the whistler frequency peaking at $\frac{1}{4}f_{ce}$. The frequency separation between the centers of stripes in the emission and absorption spectrum of the ZP structure approximates to whistler frequency (Kuijpers 1975, Chernov 1976, Maltseva \& Chernov 1989), then the frequency separation between the adjacent ZP stripes should be around two times of whistler frequency, and therefore, $\triangle f\approx\frac{1}{2}f_{ce}$ with peak whistler wave group velocity, then $M_{\alpha}\approx2$.

As for the coupling of double plasma resonance (DPR) model, the Zebra stripes are proposed to produce at some resonance levels where the upper hybrid frequency coincides with the harmonics of electron gyrofrequency in inhomogeneous magnetic flux tubes $f_{uh}=sf_{ce}$ (Berney \& Benz 1978, Winglee \& Dulk 1986, Labelle et al. 2003, Yasnov \& Karlicky 2004, Kuznetsov \& Tsap 2007). Under DPR model, the frequency separation between the adjacent ZP stripes depend on not only the magnetic field strength, but also on the scale heights of magnetic fields and plasma density in the source region (Chen et al. 2011, Karlicky \& Yasnov 2018, 2019). For example, if the emission is triggered by the coalescence of an excited plasma wave and a low frequency electrostatic wave, the emission frequency should be $f\approx f_{uh}\approx sf_{ce}$ which is called the fundamental plasma emission, and the strip frequency separation $\triangle f=\frac{H_{b}}{|sH_{b}-(s+1)H_{n}|}sf_{ce}$, $H_{b}$ and $H_{n}$ are the scale heights of magnetic field and plasma density in the source region. Obviously, here the factor $M_{\alpha}=\frac{|sH_{b}-(s+1)H_{n}|}{sH_{b}}$. When the emission is triggered by the coalescence of two plasma waves which is called as the second harmonic plasma emission, the frequency of the Zebra stripes should be $f\approx 2f_{uh}\approx sf_{ce}$ and the stripe frequency separation will be $\triangle f=\frac{H_{b}}{|sH_{b}-(s+1)H_{n}|}2sf_{ce}$, the factor $M_{\alpha}=\frac{|sH_{b}-(s+1)H_{n}|}{2sH_{b}}$. Here, we can see that the stripe frequency separation is proportional to the number of harmonics, $\triangle f\propto s$, that is, increases with emission frequency. And this is consistent with the class of growing-distant ZP (GZP) (Tan et al. 2014). Here, the scale heights of $H_{b}$ and $H_{n}$ need to be derived from other approaches, for example, the Newkirk model for plasma density (Newkirk 1961), the Dulk \& McLean model for the coronal magnetic field above the solar active region (Dulk \& McLean 1978), etc. Just because the above diagnostic functions also depends on the spatial variations of plasma density and magnetic field, this function is not an explicit function in the strict sense. Kuznetsov \& Tsap (2007) and Yasnov \& Karlicky (2020) proposed an improved
method to determine the gyro-harmonic numbers of the zebra stripes that are essential in obtaining the electron density and magnetic field strength in ZP sources.

Tan et al. (2012) applied the ZP structures to derive the magnetic fields in the source region before and after the peak time of an X2.2 flare, and the results show that the magnetic field strength is about 230-345 G in the flare impulsive phase at frequency range of 6.4-7.0 GHz, about 126-147 G after the flare peak phase at frequency range of 2.60 - 2.75 GHz, and about 23-26 G in the postflare phase at frequency range of 1.04-1.13 GHz. These diagnostic results are basically consistent with our understanding of the flaring source regions.

\textbf{Fiber Burst}

Solar radio fiber bursts have intermediate frequency drift rate which is between type II and type III bursts. They are frequently observed in solar flares (Young et al. 1961, Aurass et al. 1987, Chernov 1990,Wang et al. 2006, Wan et al. 2021). Fiber bursts may allow one to measure the strength of the coronal magnetic field in the radio burst source region. However, this is depending on the theoretical interpretations. One interpretation of the formation of fiber bursts is based upon the model of Alfven solitons which establishes the relationship between magnetic field and frequency drift rate. Therefore, radio fiber bursts can also be applied to derive the magnetic fields of the flaring source region (Kuijpers 1975, Bernold \& Treuman 1983, Wang \& Zhong 2006),

\begin{equation}
B\approx10.15\times10^{-14}H_{n}\cdot\frac{df}{dt}.
\end{equation}

$\frac{df}{dt}$ is the frequency drifting rate of the fiber burst at unit of Hz$\cdot s^{-1}$. The above function contains two variables: the scale height of plasma density ($H_{n}$) and the frequency drifting rate ($\frac{df}{dt}$). Here, $\frac{df}{dt}$ can be extracted directly from the dynamic spectrum of the source region, and $H_{n}$ should be derived indirectly from models or other approaches.

Wang \& Zhong (2006) obtained the magnetic field strength of the fiber source related to an X1.5 flare in the range of 130 - 270 G around the frequency of 2.6 - 3.8 GHz.

Another interpretation of radio fiber bursts is the whistler wave model, which presumes that the fiber bursts are caused by packets of low frequency whistler waves propagating along the magnetic field lines of a coronal loop (Benz \& Mann 1998). This model gives the magnetic field as,

\begin{equation}
B\approx17.8\times10^{-12}\frac{H_{w}}{\sqrt{x}cos\theta}\cdot\frac{df}{dt}.
\end{equation}

$H_{w}$ is the scale height in whistler wave model. $x=\frac{\omega_{w}}{\omega_{ce}}$, $\omega_{w}$ is the whistler wave frequency, $\theta$ is the propagation angle of the fiber burst and its guiding field line relative to the vertical. Generally, $x\ll0.1$. Wang \& Zhong (2006) adopted the parameter values as $x=0.01$, $H_{w}=2\times10^{5}$ km, and $\theta=0$, then obtained the magnetic field in the range 50 - 100 G for the source region of fiber bursts at frequency of 2.6-3.8 GHz, which is a bit weaker than the results derived from the above Alfven soliton model. Obviously, Equation (24) is not a complete explicit function for the parameter $x$ also containing the information of magnetic field $B$. Additionally, this equation contains several variables, $H_{w}$, $x$, $\theta$ and $\frac{df}{dt}$. Therefore, it is changeable and depending on the supplements of other methods.

\textbf{Type II Burst}

The solar radio Type II bursts are usually regarded as exciting by MHD shock waves associated with fast CMEs traveling through the solar corona with super-Alfvenic speed along the shock trajectory (Vrsnak \& Cliver 2008). They often expose the fundamental and harmonic emission band, both frequently being split in two parallel lanes that show a similar frequency drift and intensity behavior (Vrsnak et al. 2002, etc.). The frequency drifting rate of type II burst is proportional to the speed $v$ of the CME-associated shock waves and therefore linked with Alfvenic speed $v_{A}$, such relationships provides a method to estimate the magnetic field along the path of CME,

\begin{equation}
B=4\pi\frac{\sqrt{m_{e}m_{p}}}{e}\frac{H_{n}}{M_{A}}\frac{df}{dt}.
\end{equation}

Here, $M_{A}=\frac{v}{v_{A}}$ is the Alfven Mach number which could be estimated from other approaches, and generally $1.2<M_{A}<1.5$ in the regime of CMEs (Smerd et al. 1974, Vrsnak et al. 2002). $H_{n}$ is the scale height of the local plasma density which depends on certain models of plasma density, such as the Newkirk model (Newkirk 1961).

The frequency separation of splitting type II radio bursts also depends on the magnetic field strength along the path of CMEs, therefore, the splitting of type II radio bursts might be an useful tool to diagnose the magnetic fields in the related source regions. One of the interpretations of splitting type II bursts is magneto-ionic resonance which gives the frequency separation (Sturrock 1961), $\triangle f=\frac{1}{2}\frac{f_{B}^{2}}{f_{p}}$. From this we can obtain a diagnostic function of the coronal magnetic field,

\begin{equation}
B\approx5.05\times10^{-7}\sqrt{f_{p}\cdot\triangle f}.
\end{equation}

Here, the unit of $\triangle f$ and $f_{p}$ is Hz, and magnetic field $B$ is in Gauss. Many people obtained similar results from different interpretations of the formation of type II radio bursts (Vrsnak et al. 2002, Cho et al. 2007, Kishore et al. 2015, Lv et al. 2017). For example, Cho et al. (2007) obtained the coronal magnetic field strength is from 1.3 G decreasing to 0.4 G while a CME-related shock passes from 1.6 to 2.1 $R_{sun}$, and this result is agree with the potential field source surface modeling. Actually, as the type II radio bursts only occurred in the meter-wavelength and more lower frequencies ($f<300$ MHz), and Equation (20) can derive the magnetic fields only in the high corona and interplanetary space where CMEs pass through.

\textbf{Type III pair bursts}

The solar radio Type III bursts are frequently occurred on the spectrogram with fast frequency drifting rates around the period of solar flares. They are interpreted as being caused by energetic electron beams streaming through the background coronal plasma at speed of 0.1-0.9 c (c is the speed of light) (Lin \& Hudson 1971, Lin et al. 1981, Huang et al. 2007, and a review in Reid \& Ratcliffe 2014). Occasionally, we can observe type III pair bursts which are simultaneously composed of normal type III burst with negative fast frequency drifting rates and reversed slope type III bursts with positive fast frequency drifting rates (Aschwanden et al. 1993, Tan et al. 2016a). Type III pair bursts occurred mainly in the decimetric wavelengths and are interpreted as being triggered by bi-directional electron beams produced from the site of magnetic reconnection, the normal type III bursts show the propagation of upward electron beams and the reversed slope type III bursts show the downward electron beams, and therefore the separation frequency between the normal and reversed slope type III bursts might pinpoint the site of magnetic reconnection and electron accelerations. Based on the framework of plasma emission, Tan et al. (2016b) proposed a method to estimate the magnetic field around the reconnection site by using the observations of type III pair bursts,

\begin{equation}
3.40\times10^{-19}(n_{st}T\bar{D}R_{c})^{1/2}<B<3.29\times10^{-16}[\frac{n_{st}T\bar{D}R_{c}}{(n_{st}\tau)^{1/3}}]^{1/2}.
\end{equation}

In above equation, $n_{st}$ indicates the plasma density near the start site of the electron beam which can be derived from: $n_{st}=\frac{f_{st}^{2}}{81s^{2}}$. $f_{st}$ is the start frequency of the type III pair bursts. $T$ is the temperature. $\bar{D}=\frac{df}{fdt}$ is the relative frequency drifting rate. $R_{c}$ is the curvature radius of the magnetic field lines. $\tau$ is the lifetime of type III bursts. Obviously, the equation (27) can only give the range of magnetic field strength around the magnetic reconnection site. By using this method, Tan et al. (2016b) got the estimation of the magnetic field around the reconnection site in the range from about 50 G to 80 G for the X3.4 flare on 2006 December 13.

Here it should be noted that the above diagnostic functions of solar radio spectral fine structures (Equations 22 - 27) are very immature which contain many free parameters and depend on the theoretical models of the formation of the spectral fine structures. So far, the diagnostic methods of spectral fine structures can only present the magnetic field strength and even only give a wide range of the strength without information of direction of the magnetic fields. The relevant theoretical models need to be further improved and the diagnostic methods based on them also need to be optimized. Figure 1 presents a brief summary of the diagnostic functions of the magnetic field from radio observations in solar chromosphere and corona.

\begin{figure}[ht]
\begin{center}
   \includegraphics[width=14 cm]{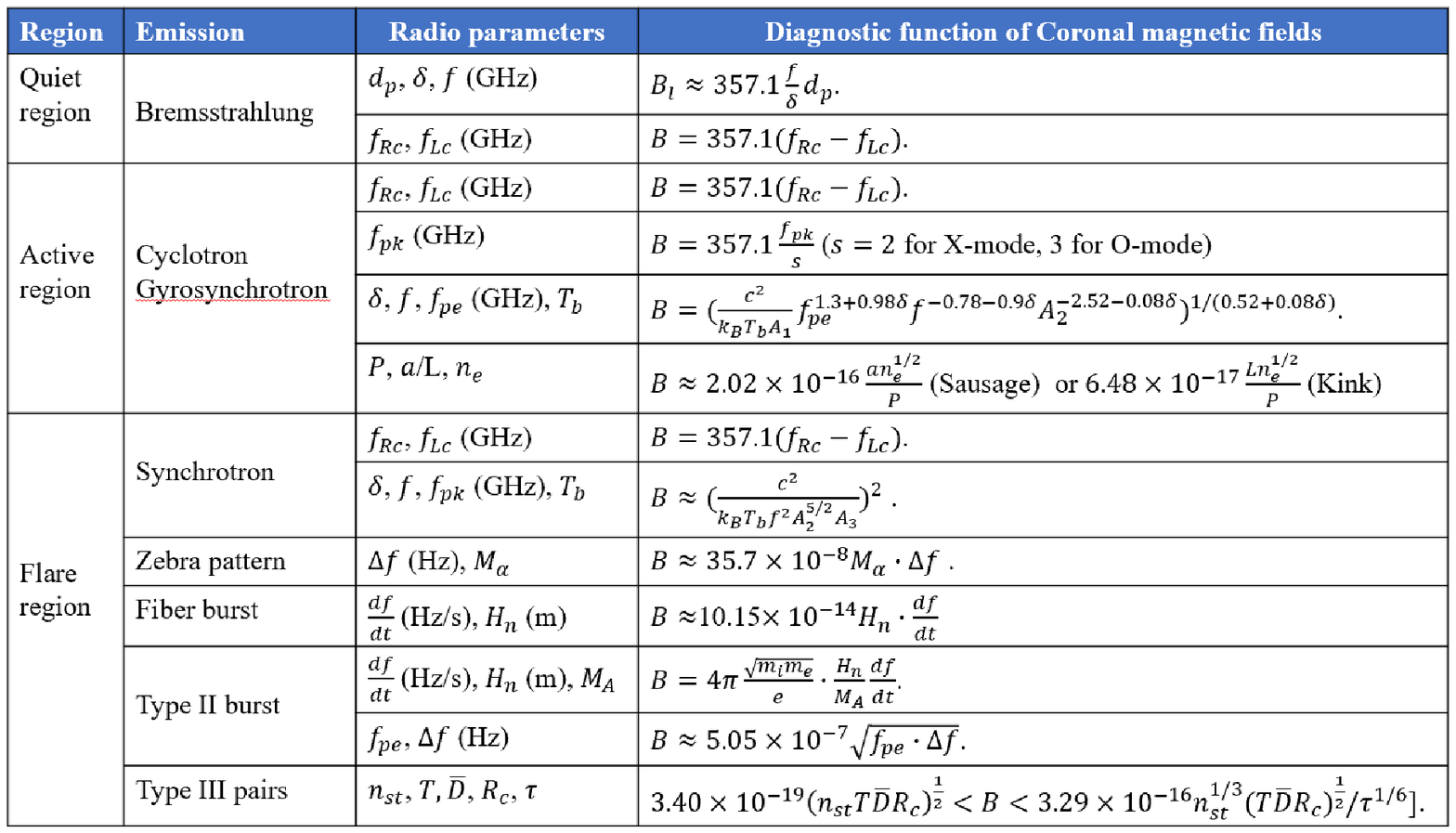}
\caption{Brief summary of the diagnostic functions of the magnetic field from radio observations in solar chromosphere and corona.}
\end{center}
\end{figure}

\section{Conclusions and discussions}

Solar radio emissions are very sensitive to magnetic fields in the very wide space of solar atmosphere, therefore, radio observations are one of the most reliable diagnostic methods of magnetic fields in the upper chromosphere and corona. This work gives 14 equations, including Equation (9), (13) and Equations (16)-(27) and forming a set of implicit diagnosing functions for deriving the chromospheric and coronal magnetic fields. Some of them do not appear in the previous reviews, such as the low-frequency cut-off diagnostics, etc. This set of diagnosing functions cover all quiet regions, active regions, and flaring source regions from the face-on to the limb of the solar full-disk. By using this set of functions and the broadband spectral solar radio imaging observations, we can derive the magnetic fields of almost any region in the solar atmosphere.

Obviously, different diagnostic functions are applicable to different physical conditions. A complete coronal magnetic map needs to consider a variety of emission mechanisms and adopt a variety of diagnostic methods. Therefore, it is necessary to partition the diagnostic area into many small diagnostic grids, and each grid has only a single emission mechanism at work. Such partitioning requires high spatial-, temporal-, and spectral-resolutions of radio observations, such as the observations obtained by MUSER (Yan et al. 2009, 2021), LOFAR (van Haarlem et al. 2013), VLA (Perley et al. 2011), and the future FASR, etc. Then we can choose an appropriate diagnostic function for each grid to derive the magnetic field strength. When we obtain the magnetic field at each grid, we can get a complete coronal magnetic map.

It is reasonable to discuss the diagnostic procedure. The diagnostics of high chromospheric and coronal magnetic fields can be divided into the several steps,

(1) Solar radio imaging observations.

It is required that the observations have high spatial-, temporal-, and spectral-resolutions, and have dual circular polarizations in a broadband frequency range, complete data calibrations (including phase and amplitude calibration, the emission intensity is in unit of sfu at each pixel), and obtain images at every frequency channel. At present, the observations can be obtained by MUSER, LOFAR, EVOSA, and even VLA are suitable for our diagnostics.

(2) Partitioning the diagnostic area.

The diagnostic source region should be divided into small grids. The scale of grids should be small as soon as possible, so that there is only one emission mechanism working in each grid. The grid partitioning also affects the accuracy of magnetic field diagnosis. Generally, the scale of grids depends on the resolutions. The higher the resolutions, the smaller the grids.

(3) Extracting observational parameters at each grid.

Generate a spectrum diagram at each grid, and extract the observational parameters for diagnostics of magnetic fields from the spectrum diagram. These observational parameters include,

a. Flux density $F$ (or brightness temperature $T_{b}$).

b. Peak frequency $f_{pk}$.

c. Cut-off frequency, $f_{Rc}$ and $f_{Lc}$.

d. Spectral index $\alpha$.

e. Degree of polarization $d_{p}$.

f. Bandwidth $\triangle f$.

g. Lifetime $\tau$.

h. Frequency drifting rate $\frac{df}{dt}$.

i. Pulsating period $P$.

j. Frequency separation $\triangle f$ for ZP.

k. Splitting frequency separation $\triangle f$ of type II burst.

And so on.

Accordingly, make the judgement of the emission mechanism at each grid. At this step, the multi-wavelength EUV images with high spatial resolutions (such as AIA images) and the vector magnetograph (such as HMI images) are the very good supplementary materials which may help us to identify the mechanism.

(4) Deriving magnetic fields.

After we obtained all the observational parameters and identified the emission mechanism at each grids, we may derive the magnetic fields for every grid. When the grid is located in the quiet region, the magnetic field can be calculated from Equation (13) or (16). When the grid is located in the active region, the magnetic field can be calculated from Equations (17) and (18). At the same time, if we can extract the signal of QPP, we can also use the Equation (19) and (20) to give the magnetic field. When the grid is located in the flaring source region during the nonthermal eruptions, the magnetic field can be calculated from Equation (21), and at the same time, if we can identify the kinds of spectral fine structures (such as ZPs, fiber bursts, splitting type II bursts, etc.) related to the grid, we may derive the magnetic field from Equation (22), or Equation (23), or Equation (24).

After we derived the magnetic fields at each grid, then a full coronal magnetic field map will be completed. Here, we meet another question: Are these diagnostic results credible? We may apply several methods to verify these results.

The first simple method is to compare the diagnosing result with the ones deriving from a simple modeling function proposed by Dulk \& McLean (1978),

\begin{equation}
B\approx0.5(\frac{R_{sun}}{h})^{3/2}.
\end{equation}

The unit of $B$ is G, $R_{sun}$ is the solar photospheric radius and $h$ is the height of the source region above the solar photosphere. The comparison between the diagnosing result and the calculating ones from Equation (26) can preliminarily verify the reliability of the diagnosis results. In the work of Tan et al. (2012), the diagnosing results of microwave ZPs were verified by above methods and they are consistent with each other.

Another method is to compare the diagnosing result with MHD modeling and extrapolations (such as Yan \& Sakurai 2000, Yan \& Li 2006, Metcalf et al. 2008, Wiegelmann 2008, Jiang \& Feng 2013, Chifu et al. 2015, Zhu \& Wiegelman 2018, etc.) from direct measurements in the photosphere and, to a lesser degree, in the chromosphere, which are based mainly on the Zeeman effect. Such comparison can not only verify the reliability of the diagnosis results, but also optimize and improve the modeling extrapolation methods.

When the source region locates at the solar limb region, we may apply the simultaneous observations of radio imaging observations and Zeeman splitting observations of infrared coronal forbidden emission lines (Lin et al. 2004, Liu \& Lin 2008) and the results of coronal MHD waves diagnostics (Nakariakov \& Verwichte 2005, Yang et al. 2020) to cross verify the above diagnosing results.

Generally speaking, the coronal magnetic field diagnostics is a key problem in solar physics. However, the diagnosing functions presented in this work are approximation at present, some of them have limitations, for example, so far we have no effective method to obtain the transverse magnetic field component of the quiet region, and the diagnostics of spectral fine structures can not provide the information of direction of magnetic fields. There are still many aspects of these functions that deserve modification and improvement in the future works. In the next works, we will attempt to apply the MUSER solar radio imaging observations and the above diagnosing functions to produce the solar full-disk coronal magnetic maps in different frequencies, which may help us to understand a series of important problems of solar physics, such as the mechanism of coronal heating, the origin of solar flares and their precursors, and their possible relationship with space weather, etc.

\begin{acknowledgements}
This work is supported by NSFC Grants 11790301, 11973057, 12003048 and 11941003, National Key R\&D Program of China 2021YFA1600500, 2021YFA1600503, and the International Partnership Program of Chinese Academy of Sciences (183311KYSB20200003).
\end{acknowledgements}

\label{lastpage}


\begin{thebibliography}{99}
\bibitem[Abrami(1970)]{Abrami1970} Abrami, A. 1970, Sol Phys, 11, 104.

\bibitem[Akhmedov(1982)]{Akhmedov1982} Akhmedov, S.B., Gelfreikh, G.B., Bogod, V.M., Korzhavin, A.N. 1982, Sol Phys, 79, 41.

\bibitem[Alissandrakis(1980)]{Alissandrakis1980} Alissandrakis, C.E., Kundu, M.R., Lantos, P. 1980, A\&A, 82, 30.

\bibitem[Alissandrakis(2019)]{Alissandrakis2019} Alissandrakis, C.E., Bogod, V.M., Kaltman, T.I., Patsourakos, S., Peterova, N.G. 2019, Sol Phys, 294, 23.

\bibitem[Alissandrakis(2021)]{Alissandrakis2021} Alissandrakis, C.E., Gary, D.E. 2021, Frontiers in Astron Space Sci., 7, 591075.

\bibitem[Altyntsev(2005)]{Altyntsev2005} Altyntsev, A.T., Kuznetsov, A.A., Meshalkina, N.S., Rudenko, G.V., Yan, Y. 2005, A\&A, 431, 1037.

\bibitem[Andries(2009)]{Andries2009} Andries, J., van Doorsselaere, T., Roberts, B., Verth, G., Verwichte, E., Erdelyi, R. 2005, A\&A, 431, 1037.

\bibitem[Appleton(1932)]{Appleton1932} Appleton, E.V. 1932, J. Inst. Electr. Eng., 71, 642.

\bibitem[Aschwanden(1987)]{Aschwanden1987} Aschwanden, M.J. 1987, Sol Phys, 111, 113.

\bibitem[Aschwanden(1993)]{Aschwanden1993} Aschwanden, M.J., Benz, A.O., Schwartz, R.A. 1993, ApJ, 417, 790.

\bibitem[Aurass(1987)]{Aurass1987} Aurass, H., Chernov, G.P., Karlicky, M., Kurths, J., Mann, G. 1987, Sol Phys, 112, 347.

\bibitem[Bastian(1998)]{Bastian1998} Bastian, T.S., Benz, A.O., Gary, D.E. 1998, Annu Rev Astron Astrophys, 36, 131.

\bibitem[Beiersdorfer(2003)]{Beiersdorfer2003} Beiersdorfer, P., Scofield, J. H., Osterheld, A. L. 2003, Phys Rev Lett, 90, 235003.

\bibitem[Benz(1998)]{Benz1998} Benz, A.O., Mann, G. 1998, A\&A, 333, 1034.

\bibitem[Benz(2009)]{Benz2009} Benz, A.O. 2009, Solar System, Landolt-Bornstein - Group VI Astronomy and Astrophysics, Springer-Verlag Berlin Heidelberg, 4B, 103.

\bibitem[Berney(1978)]{Berney1978} Berney, M., Benz, A.O. 1978, A\&A, 65, 369.

\bibitem[Bernold(1983)]{Bernold1983} Bernold, T.E.X., Treuman, R.A. 1983, ApJ, 264, 677.

\bibitem[Bogod(1980)]{Bogod1980} Bogod, V.M., Gelfreikh, G.B. 1980, Sol Phys, 67, 29.

\bibitem[Casini(2017)]{Casini2017} Casini, R., White, S.M., Judge, P.G. 2017, Space Sci Rev, 210, 145.

\bibitem[Chen(2011)]{Chen2011} Chen, B., Bastian, T. S., Gary, D. E., Jing, J. 2011, ApJ, 736, 64.

\bibitem[Chen(2020)]{Chen2020} Chen, B., Shen, C.C., Gary, D.E., Reeves, K.K., Fleishman, G.D., Yu, S.J., et al. 2020, Nature Astron., 4, 1140.

\bibitem[Chen(2021)]{Chen2021} Chen, Y.J., Liu, X.Y., Tian, H., Bai, X.Y., Jin, M., Li, W.X., et al. 2021, ApJ Lett, 918, 13.

\bibitem[Chernov(1976)]{Chernov1976} Chernov, G. 1976, Soviet Astron., 20, 582.

\bibitem[Chernov(1990)]{Chernov1990} Chernov, G. 1990, Soviet Astron., 34, 66.

\bibitem[Chernov(2003)]{Chernov2003} Chernov, G., Yan, Y.H., Fu, Q.J. 2003, A\&A, 406, 1071.

\bibitem[Chernov(2006)]{Chernov2006} Chernov, G. 2006, Space Sci. Rev., 127, 195.

\bibitem[Chernov(2010)]{Chernov2010}Chernov, G.P. 2010, \emph{RAA} \textbf{10}, 821.

\bibitem[Chernov et al(2010)]{Chernov et al2010}Chernov, G.P., Yan, Y.H., Tan, C.M., Chen, B., Fu, Q.J. 2010, \emph{Sol Phys} \textbf{262}, 149.

\bibitem[Chernov(2011)]{Chernov2011} Chernov, G. 2011, Fine Structure of Solar Radio Burst: Berlin: Springer.

\bibitem[Chifu(2015)]{Chifu2015} Chifu, I., Inhester, B., Wiegelmann, T. 2015, A\&A, 577, 123.

\bibitem[Cho(2007)]{Cho2007} Cho, K.S., Lee, J., Gary, D.E., Moon, Y.J., Park, Y.D. 2007, ApJ, 665, 799.

\bibitem[Davila(1987)]{Davila1987} Davila, J. 1987, ApJ, 317, 514.

\bibitem[De Pontieu(2007)]{De Pontieu2007} De Pontieu, B., McIntosh, S.W., Carlsson, M., et al. 2007, Science, 318, 574.

\bibitem[De Moortel(2009)]{De Moortel2009} De Moortel, I., Pascoe, D. J. 2009, ApJ, 669, 72.

\bibitem[Dulk(1978)]{Dulk1978} Dulk, G.A., McLean, D.J. 1978, Sol Phys, 57, 279.

\bibitem[Dulk(1982)]{Dulk1982} Dulk, G.A., Marsh, K.A. 1982, ApJ, 259, 350.

\bibitem[Dulk(1985)]{Dulk1985} Dulk, G.A. 1985, Annu Rev Astron Astrophys, 23, 169.

\bibitem[Elgaroy(1959)]{Elgaroy1959} Elgaroy, O. 1959, Nature, 184, 887.

\bibitem[Fleishman(2003)]{Fleishman2003} Fleishman, G.D., Melnikov, V.F. 2003, ApJ, 587, 823.

\bibitem[Fleishman(2020)]{Fleishman2020} Fleishman, G.D., Gary, D.E., Chen, B., Kuroda, N., Yu, S.J., Nita, G.M. 2020, Science, 367, 278.

\bibitem[Fletcher(2011)]{Fletcher2011} Fletcher, L., Dennis, B.R., Hudson, H.S., et al. 2011, Space Sci Rev, 159, 19.

\bibitem[Gary(2008)]{Gary2008} Gary, D.E., Chen, B., Dennis, B.R., Fleishman, G.D., Hurford, G.J., Krucker, S., et al. 2008, ApJ, 863, 38.

\bibitem[Gelfreikh(1987)]{Gelfreikh1987} Gelfreikh, G.B., Peterova, N. G., Riabov, B. I. 1987, Sol. Phys., 108, 89.

\bibitem[Ginzburg(1958)]{Ginzburg1958} Ginzburg, V.L., Zheleznyakov, V.V. 1958, Astron Zh., 35, 694.

\bibitem[Hale(1908)]{Hale1908} Hale, G.E. 1908, PASP, 20, 287.

\bibitem[Hanle(1924)]{Hanle1924} Hanle, W. 1924, Zeitschrift fur Physik, 30, 93.

\bibitem[Hartree(1931)]{Hartree1931} Hartree, D.R. 1931, Proc. Cambridge Philos. Soc., 27, 143.

\bibitem[Hatanaka(1956)]{Hatanaka1956} Hatanaka, W. 1956, PASJ, 8, 73.

\bibitem[Huang(2002)]{Huang2002} Huang, G.L., Nakajima, H. 2002, New Astron, 7, 135.

\bibitem[Huang(2006)]{Huang2006} Huang, G.L. 2006, Sol Phys, 237, 173.

\bibitem[Huang(2008)]{Huang2008} Huang, G.L. 2008, Adv Space Res., 41, 1191.

\bibitem[Huang(2013)]{Huang2013} Huang, G.L., Li, J.P., Song, Q.W. 2013, Res Astron Astrophys, 13, 215.

\bibitem[Huang(2007)]{Huang2007} Huang, J., Yan, Y.H., Liu, Y.Y. 2007, Adv Space Res, 39, 1439.

\bibitem[Hudson(2011)]{Hudson2011} Hudson, H.S. 2011, Space Sci Rev, 158, 5.

\bibitem[Ignace(1997)]{Ignace1997} Ignace, R., Nordsieck, K.H., Cassinelli, J.P. 1997, ApJ, 486, 550.

\bibitem[Inglis(2016)]{Inglis2016} Inglis, A. R., Ireland, J., Dennis, B. R., Hayes, L., Gallagher, P. 2016, ApJ, 833, 284.

\bibitem[Jess(2016)]{Jess2016} Jess, D.B., Reznikova, V.E., Ryans, R.S.I., Christian, D., J., Keys, P.H., Mathioudakis, M., Et al. 2016, Nature Phys, 12, 179.

\bibitem[Jiang(2013)]{Jiang2013} Jiang, C.W., Feng, X.S. 2013, ApJ, 769, 144.

\bibitem[Judge(2013)]{Judge2013} Judge, P.G., Habbal, S., Landi, E. 2013, Sol Phys, 288, 467.

\bibitem[Kakinuma(1962)]{Kakinuma1962} Kakinuma, T., Swarup, G., 1962, ApJ, 136, 975

\bibitem[Karlicky(2013)]{Karlicky2013}Karlicky, M., Meszarosova, H., Jelinek, P. 2013, \emph{Astron. Astrophys.} \textbf{550}, 1.

\bibitem[Karlicky(2018)]{Karlicky2018}Karlicky, M., Yasnov, L.V. 2018, \emph{Astron. Astrophys.} \textbf{618}, 60.

\bibitem[Karlicky(2019)]{Karlicky2018}Karlicky, M., Yasnov, L.V. 2019, \emph{Astron. Astrophys.} \textbf{624}, 119.

\bibitem[Kishore(2015)]{Kishore2015} Kishore, P., Ramesh, R., Hariharan, K., Kathiravan, C., Goplaswamy, N., 2015, ApJ, 832, 59

\bibitem[Kruger(1993)]{Kruger1993} Kruger, A., Hildebrandt, J., 1993, ASPC, 46, 294

\bibitem[Kuijpers(1975)]{Kuijpers1975} Kuijpers, J. 1975, Sol Physt 44, 173

\bibitem[Kundu(1990)]{Kundu1990} Kundu, M.R. 1990, Mem SA It 61, 431

\bibitem[Kuznetsov(2007)]{Kuznetsov2007} Kuznetsov, A.A., Tsap, Yu. T., 2007, Sol Phys, 241, 127.

\bibitem[LaBelle(2003)]{LaBelle2003} LaBelle, J., Treumann, R.A., Yoon, P.H., Karlicky, M. 2003, ApJ, 593, 1195.

\bibitem[Lee(2007)]{Lee2007} Lee, J. 2007, Space Sci Rev, 133, 73.

\bibitem[Li(2015)]{Li2015} Li, D., Ning, Z. J., Zhang, Q. M. 2015, ApJ, 807, 72.

\bibitem[Li(2015)]{Li2015} Li, W., Grumer, J., Yang, Y., Brage, T, Yao, K., Chen, C.Y., Watanabe, T., Jonsson, P., Lundstedt, H., Hutton, R., Zou, T.M. 2015, ApJ, 807, 69.

\bibitem[Li(2021)]{Li2021} Li, W., Li, M., Wang, K., Brage, T., Hutton, R., Landi, E. 2021, ApJ, 913, 135.

\bibitem[Lin(1971)]{Lin1971} Lin, R.P., Hudson, H.S. 1971, Sol Phys, 17, 412.

\bibitem[Lin(1981)]{Lin1981} Lin, R.P., Potter, D.W., Gurnett, D.A., Scarf, F.L. 1981, ApJ, 251, 364.

\bibitem[Lin(2004)]{Lin2004} Lin, H., Kuhn, J.R., Coulter, R. 2004, ApJ Letter, 613, L177.

\bibitem[Liu(2008)]{Liu2008} Liu, Y., Lin, H.S. 2008, ApJ, 680, 1496.

\bibitem[Lv(2017)]{Lv2017} Lv, M.S., Chen, Y., Li, C.Y., Zimovets, I., Du, G.H., Wang, B., Feng, S.W., Ma, S.L. 2017, Sol Phys, 292, 194.

\bibitem[Metcalf(2008)]{Metcalf2008} Metcalf, T., De Rosa, M.L., Schrijver, C.A., Barnes, G., van Ballegooijen, A.A., Wiegelmann, T. 2008, Sol Phys, 247, 269.

\bibitem[Maltseva(1989)]{Maltseva1989} Maltseva, O.A., Chernov, G.P., 1989, Kinematika i Fizika Nebesnykh Tel, 5, 44.

\bibitem[Melrose(1980)]{Melrose1980} Melrose, D.B. 1980, Space Sci Rev, 26, 3.

\bibitem[Nakariakov(1999)]{Nakariakov1999} Nakariakov, V. M., Ofman, L., Deluca, E. E., Roberts, B., Davila, J. M. 1999, Science, 285, 862.

\bibitem[Nakariakov(2001)]{Nakariakov2001} Nakariakov, V. M., Ofman, L. 2001, ApJ, 372, L53.

\bibitem[Nakariakov(2005)]{Nakariakov2005} Nakariakov, V.M., Verwichte, E. 2005, Living Rev Sol Phys, 2, 3.

\bibitem[Nakariakov(2009)]{Nakariakov2009} Nakariakov, V.M., Melnikov, V.F. 2009, Space Sci Rev, 149, 119.

\bibitem[Nakariakov(2010)]{Nakariakov2010} Nakariakov, V.M., Foullon, C., Myagkova, I.N., Inglis, A.R 2010, ApJ, 708, L47.

\bibitem[Newkirk(1961)]{Newkirk1961} Newkirk, G.J. 1961, AJ, 133, 983.

\bibitem[Nindos(2008)]{Nindos2008} Nindos, A., Aurass, H., Klein, K.L., Trottet, G. 2008, Sol Phys, 253, 3.

\bibitem[Nindos(2019)]{Nindos2019} Nindos, A., Kontar, E. P., Oberoi, D. 2019, Adv Space Res, 63, 1404.

\bibitem[Nindos(2020)]{Nindos2020} Nindos, A., 2019, Front. Astron. Space Sci. 7, 57

\bibitem[Orozco Suarez(2007)]{Orozco Suarez2007} Orozco Suarez, D., Bellot Rubio, L.R., del Toro Iniesta, J.C., Tsuneta, S., Lites, B.W., Ichimoto, K., et al. 2007, ApJ, 670, L61

\bibitem[Parker(1988)]{Parker1988} Parker, E.N. 1988, ApJ, 330, 474

\bibitem[Perley(2011)]{Perley2011} Perley, R.A., Chandler, C.J., Butler, B.J., Wrobel, J.M. 2011, ApJ, 739, L1

\bibitem[Rappazzo(2008)]{Rappazzo2008} Rappazzo, A.F., Velli, M., Einaudi, G., \& Dahlburg, R.B. 2008, ApJ, 677, 1348

\bibitem[Raulin(2005)]{Raulin2005} Raulin, J.P., Pacini, A.A. 2005, Adv Space Res, 35, 739

\bibitem[Peid(2014)]{Reid2014} Reid, H.A.S., Ratcliffe, H. 2014, RAA, 14, 773

\bibitem[Raouafi(2016)]{Raouafi2016} Raouafi, N.E., Riley, P., Gibson, S., Fineschi, S., Solanki, S. K. 2016, Frontiers in Astron Space Sci., 3, 20.

\bibitem[Ratcliffe(1959)]{Ratcliffe1959} Ratcliffe, J.A., 1959, The magneto-Ionic Theory and Its Applications to the Ionosphere, Cambridge, Cambridge University Press.

\bibitem[Robinson(1984)]{Robinson1984} Robinson, P.A., Melrose, D.B. 1984, Aust J. Phys, 37, 675.

\bibitem[Robinson(2000)]{Robinson2000} Robinson, P.A., Benz, A.O. 2000, Sol Phys, 194, 345.

\bibitem[Rosenberg(1972)]{Rosenberg1972} Rosenberg, H. 1972, Sol Phys, 25, 188.

\bibitem[Spangler(2005)]{Spangler2005} Spangler, S.R. 2005, Space Sci Rev, 121, 189

\bibitem[Schneider(1959)]{Schneider1959} Schneider, J. 1959, Phys. Rev. Lett., 2, 504

\bibitem[Seehafer(1986)]{Seehafer1986} Seehafer, N. 1986, Sol Phys, 105, 223

\bibitem[Shibasaki(2011)]{Shibasaki2011} Shibasaki, K., Alissandrakis, C. E., Pohjolainen, S. 2011, Sol Phys, 273, 309

\bibitem[Slottje(1972)]{Slottje1972} Slottje, C. 1972, Sol Phys, 25, 210

\bibitem[Smerd(1974)]{Smerd1974} Smerd, S.F., Sheridan, K.V., Stewart, R.T. 1974, IAU Symposium, 57, 389

\bibitem[Sturrock(1961)]{Sturrock1961} Sturrock, P.A. 1961, Nature, 192, 58

\bibitem[Sturrock(1999)]{Sturrock1999} Sturrock, P.A. 1999, ApJ, 521, 451

\bibitem[Tan(2010)]{Tan2010} Tan, B.L., Zhang, Y., Tan, C.M., Liu, Y.Y. 2010, ApJ, 723, 25.

\bibitem[Tan(2012)]{Tan2012} Tan, B.L., Yan, Y.H, Tan, C.M., Sych, R., Gao, G.N. 2012, ApJ, 744, 166.

\bibitem[Tan(2013)]{Tan2013} Tan, B.L. 2013, \emph{ApJ}, \textbf{773}, 165.

\bibitem[Tan(2014)]{Tan2014} Tan, B.L., 2014, ApJ, 795, 140.

\bibitem[Tan(2014)]{Tan2014} Tan, B.L., Tan, C.M., Zhang, Y., Meszarosova, H., Karlicky, M. 2014, ApJ, 780, 129.

\bibitem[Tan(2016a)]{Tan2016a} Tan, B.L., Meszarosova, H., Karlicky, M., Huang, G.L., Tan, C.M. 2016a, ApJ, 819, 42.

\bibitem[Tan(2016b)]{Tan2016b} Tan, B.L., Karlicky, M., Meszarosova, H., Huang, G.L., 2016b, RAA, 16, 82.

\bibitem[Tan(2020)]{Tan2020} Tan, B.L., Yan, Y., Li, T., Zhang, Y., Chen, X.Y., 2020, RAA, 20, 90.

\bibitem[Tan(2021)]{Tan2021} Tan, B.L., 2021, Universe, 7, 378.

\bibitem[Tang(2016)]{Tang2016} Tang, J.F., Wu, D.J., Chen, L., Zhao, G. Q., Tan, C. M. 2016, ApJ, 823, 8.

\bibitem[Tang(2021)]{Tang2021} Tang, J.F., Wu, D.J., Wan J.L., Chen, L., Tan, C.M. 2021, \emph{RAA}, \textbf{21}, 148.

\bibitem[Twiss(1958)]{Twiss1958} Twiss, R.O., 1958, Aust. J. Phys., 11, 564.

\bibitem[van Haarlem(2013)]{van Haarlem2013} van Haarlem, M.P., Wise, M.W., Gunst, A.W., et al. 2013, A\&A, 556, A2.

\bibitem[Wan(2021)]{Wan2021} Wan, J.L., Tang, J.F., Tan, B.L., Shen, J.H., Tan, C.M. 2021, A\&A, 653, A38.

\bibitem[Wang(1995)]{Wang1995} Wang, J.X., Wang, H.M, Tang, F., Lee, J.W., Zirin, H. 1995, Sol Phys, 160, 277.

\bibitem[Wang(2006)]{Wang2006} Wang, S.J., Zhong, X.C. 2006, Sol Phys, 236, 155.

\bibitem[Wang(2008)]{Wang2008} Wang, S.J., Yan, Y.H., Liu, Y.Y., Fu, Q.J., Tan, B.L., Zhang, Y. 2008, Sol Phys, 253, 133.

\bibitem[Wang(2015)]{Wang2015} Wang, Z.T., Gary, D.E., Fleishman, G.D., White, S.M. 2015, ApJ, 805, 93.

\bibitem[White(2005)]{White2005} White, S.M. 2005, ESASP, 596, E10.

\bibitem[Wiegelmann(2008)]{Wiegelmann2008} Wiegelmann, T. 2008, JGRA, 113, A03S02.

\bibitem[Wiegelmann(2014)]{Wiegelmann2014} Wiegelmann, T., Thalmann, J.K., Solanki, S.K. 2014, Astron Astrophys Rev, 22, 78.

\bibitem[Winglee(1986)]{Winglee1986} Winglee, R.M., Dulk, G.A. 1986, ApJ, 307, 808.

\bibitem[Wu(1979)]{Wu1979} Wu, C.S., Lee, L.C. 1979, ApJ, 230, 621.

\bibitem[Vrsnak(2002)]{Vrsnak2002} Vrsnak, B., Magdalalenic, J., Aurass, H., Mann, G. 2002, A\&A, 396, 673.

\bibitem[Vrsnak(2008)]{Vrsnak2008} Vrsnak, B., Cliver, E.W. 2008, Sol Phys, 253, 215.

\bibitem[Yan(2000)]{Yan2000} Yan, Y.H., Sakurai, T., 2000, Sol. Phys, 195, 89.

\bibitem[Yan(2006)]{Yan2006} Yan, Y.H., Li, Z.H, 2006, ApJ, 638, 1162.

\bibitem[Yan(2009)]{Yan2009} Yan, Y.H., Zhang, J., Wang, W., et al. 2009, Earth Moon Planet, 104, 97.

\bibitem[Yan(2020)]{Yan2020} Yan, Y.H., Tan B.L., Melnikov, V., Chen, X.Y., Wang, W., Chen, L.J., et al. 2020, IAUS, 354, 17.

\bibitem[Yan(2021)]{Yan2021} Yan, Y.H., Chen, Z.J, Wang, W., Liu, F., Geng, L.H., Chen, L.J., et al. 2021, Frontiers Astron Space Sci, 8, 584043.

\bibitem[Yang(2020)]{Yang2020} Yang, Z.H., Bethge, C., Tian, H., Tomczyk, S., Morton, R., Del Zanna, G., et al. 2020, Science, 369, 694.

\bibitem[Yasnov(2004)]{Yasnov2004} Yasnov, L.V., Karlicky, M. 2004, Sol Phys, 219, 289.

\bibitem[Yasnov(2020)]{Yasnov2020} Yasnov, L.V., Karlicky, M. 2020, Sol Phys, 295, 96.

\bibitem[Young(1961)]{Young1961} Young, C.W., Spencer, C.L., Moreton, G.E., Roberts, J.A. 1961, ApJ, 133, 243.

\bibitem[Zaitsev(1983)]{Zaitsev1983} Zaitsev, V.V., Stepanov, A.V., 1983, Sol Phys, 88, 297.

\bibitem[Zheleznyakov(1975)]{Zheleznyakov1975} Zheleznyakov, V.V., Zlotnik, E.Y., 1975, Sol Phys, 44, 461.

\bibitem[Zhou(1994)]{Zhou1994} Zhou, A.H., Karlicky, M. 1994, Sol Phys, 153, 441.

\bibitem[Zhou(2013)]{Zhou2013} Zhou, G.P., Wang, J.X., Jin, C.L. 2013, Sol Phys, 283, 273.

\bibitem[Zhu(2021)]{Zhu2021} Zhu, R., Tan, B.L., Su, Y.N., Tian, H., Xu, Y., Chen, X.Y., Song, Y.L., Tan, G.Y., 2021, Sci China Tech Sci,, 64, 169.

\bibitem[Zhu(2018)]{Zhu2018} Zhu, X.S., Wiegelmann, T. 2018, ApJ, 866, 130.

\bibitem[Zimovets(2021)]{Zimovets2021} Zimovets, I.V., McLaughlin, J.A., Srivastava, A.K., Kolotkov, D.Y., Kuznetsov, A.A., Kupriyanova, E.G., et al. 2021, Space Sci Rev, 217, 66.

\end{thebibliography}
\end{document}